\documentclass[aps,prl,final,twocolumn,letterpaper]{revtex4}

\usepackage{graphicx}   
\usepackage{import}                         % Graphics package
\usepackage{epstopdf}
\usepackage{amsmath} 
\usepackage{float} 
\usepackage{bm}
\usepackage{amssymb}
\usepackage{quotes}
\usepackage{indentfirst}
\usepackage{color}
\usepackage{transparent}
\usepackage{dcolumn}% Align table columns on decimal point
\usepackage{braket}
\usepackage{multirow}
\usepackage{cancel} % for the command \cancel and \cancelto
\usepackage{mdframed}
\usepackage{color}
\usepackage{bm}
\usepackage{dsfont}
\usepackage{slashed}
\usepackage{graphicx}
\usepackage{amsmath}
\usepackage{siunitx}
\usepackage{hyperref}
\usepackage{soul, color} % for the \hl and \st commands
\soulregister\ref{7}  % so that \hl and \st can wrap around \rf
\soulregister\cite{7} % so that \hl and \st can wrap around \cite
\renewcommand{\st}[1]{}

\usepackage{xr}

\makeatletter
\newcommand*{\addFileDependency}[1]{% argument=file name and extension
  \typeout{(#1)}
  \@addtofilelist{#1}
  \IfFileExists{#1}{}{\typeout{No file #1.}}
}
\makeatother

\usepackage{textcomp} % for \textrightarrow
\usepackage{xifthen}
\usepackage{xcolor}
\usepackage{etoolbox}
\newboolean{togglechanges} 

\setboolean{togglechanges}{false}

\newcommand{\comment}[1]{\ifbool{togglechanges}
    {#1}  % updates-only version
    {\textcolor{blue}{#1}}}

\usepackage{bibentry}

\usepackage{graphicx}% Include figure files
\usepackage{dcolumn}% Align table columns on decimal point
\usepackage{bm}% bold math

\usepackage{textcomp} % for \textrightarrow

\begin{document}
\rmfamily

\title{Large-scale self-assembled nanophotonic scintillators for X-ray imaging}

\author{Louis~Martin-Monier$^{\P,1}$}
\email{lmmartin@mit.edu}
\author{Simo~Pajovic$^{\P,2}$}
\email{pajovics@mit.edu}
\author{Muluneh~G.~Abebe$^{\P,3,4}$}
\email{abebe181@mit.edu}
\author{Joshua~Chen$^{5}$}
\author{Sachin~Vaidya$^{3}$}
\author{Seokhwan~Min$^{3,6}$}
\author{Seou~Choi$^{5}$}
\author{Steven~E.~Kooi$^{7}$}
\author{Bjorn~Maes$^{4}$}
\author{Juejun~Hu$^{1}$} 
\author{Marin~Soljačić$^{3,5}$}
\author{Charles~Roques-Carmes$^{5,8}$}
\email{chrc@stanford.edu}

\affiliation{$^{1}$ Department of Materials Science and Engineering, MIT, Cambridge MA USA \looseness=-1}
\affiliation{$^{2}$ Department of Mechanical Engineering, MIT, Cambridge MA USA\looseness=-1}
\affiliation{$^{3}$ Department of Physics, Massachusetts Institute of Technology, Cambridge, Massachusetts 02139, USA\looseness=-1}
\affiliation{$^{4}$ Micro- and Nanophotonic Materials Group, Research Institute for Materials Science and Engineering, University of Mons, 20 Place du Parc, Mons B-7000, Belgium\looseness=-1}
\affiliation{$^{5}$ Research Laboratory of Electronics, MIT, Cambridge MA USA\looseness=-1}
\affiliation{$^{6}$ Department of Materials Science and Engineering, Korea Advanced Institute of Science and Technology, Daejeon 34141, Republic of Korea\looseness=-1}
\affiliation{$^{7}$ Institute for Soldier Nanotechnologies, MIT, Cambridge MA USA\looseness=-1}
\affiliation{$^{8}$ E. L. Ginzton Laboratory, Stanford University, 348 Via Pueblo, Stanford, CA 94305\looseness=-1
\\
$^{\P}$ denotes equal contribution.}

% \noindent	

% \noindent

\clearpage 

%-----CHANGE SETUP FOR PARAGRAPH INDENTS AND SKIPS-----
% \setlength{\parindent}{0em}
% \setlength{\parskip}{.5em}
\vspace*{-2em}

%-------------------------------------
%------------- MAIN TEXT -------------
%-------------------------------------

%%%%%%%%%%%%%%%%%%%%%%%%%%%%%%%%%%%%%%%%%%%%%%%%%%%%%%%%%%%%%%%%%%%%%%%%%%%%%%%%%%%%%%%%%%%%%%%%%%%%%%%%%
%%% INSTRUCTIONS FOR COMMENTS AND MODIFICATIONS -- PLEASE RESPECT THEM, WILL SAVE A LOT OF TIME 
% - Make your comments either using Review tools or \comment{}
% - To refer to Figures: Figure~\ref{fig:X}a
% - Same for Equations 
% - To Refer to SI sections, do manually for now 
%%%%%%%%%%%%%%%%%%%%%%%%%%%%%%%%%%%%%%%%%%%%%%%%%%%%%%%%%%%%%%%%%%%%%%%%%%%%%%%%%%%%%%%%%%%%%%%%%%%%%%%%%

\begin{abstract}
Scintillators are essential for converting X-ray energy into visible light in imaging technologies. Their widespread application in imaging technologies has been enabled by scalable, high-quality, and affordable manufacturing methods. Nanophotonic scintillators, which feature nanostructures at the scale of their emission wavelength, provide a promising approach to enhance emission properties like light yield, decay time, and directionality. However, scalable fabrication of such nanostructured scintillators has been a significant challenge, impeding their widespread adoption. Here, we present a scalable fabrication method for large-area nanophotonic scintillators based on the self-assembly of chalcogenide glass photonic crystals. This technique enables the production of nanophotonic scintillators over wafer-scale areas, achieving a six-fold enhancement in light yield compared to unpatterned scintillators. We demonstrate this approach using a conventional X-ray scintillator material, cerium-doped yttrium aluminum garnet (YAG:Ce). By analyzing the influence of surface nanofabrication disorder, we establish its effect on imaging performance and provide a route towards large-scale scintillation enhancements without decrease in spatial resolution. Finally, we demonstrate the practical applicability of our nanophotonic scintillators through X-ray imaging of biological and inorganic specimens. Our results indicate that this scalable fabrication technique could enable the industrial implementation of a new generation of nanophotonic-enhanced scintillators, with significant implications for advancements in medical imaging, security screening, and nondestructive testing.\end{abstract}

\maketitle

\section{Introduction}

Scintillation, the process by which materials emit light upon exposure to high-energy particles such as X-rays, is paramount to numerous technologies~\cite{gektin2017inorganic}. Its role is particularly prominent in X-ray imaging and characterization, where scintillators are crucial for converting X-ray energy into visible light, which can then be detected and analyzed~\cite{nagarkar2004new}. Advances in bulk scintillator processing have been key to their widespread adoption in X-ray imaging applications. Techniques like the Czochralski and Bridgman methods~\cite{yoshikawa2013czochralski, boatner2014advances} are highly scalable and have been adapted  to produce large, high-quality scintillator crystals in bulk. Other established techniques such as thermal evaporation~\cite{cha2010fabrication, bhandari2012hot} and sol-gel methods~\cite{nedelec2007sol} have also been successfully tailored for large-area polycrystalline scintillator manufacturing.

\begin{figure*}
\centering
\vspace{-0.2cm}
  \includegraphics[scale=0.65]{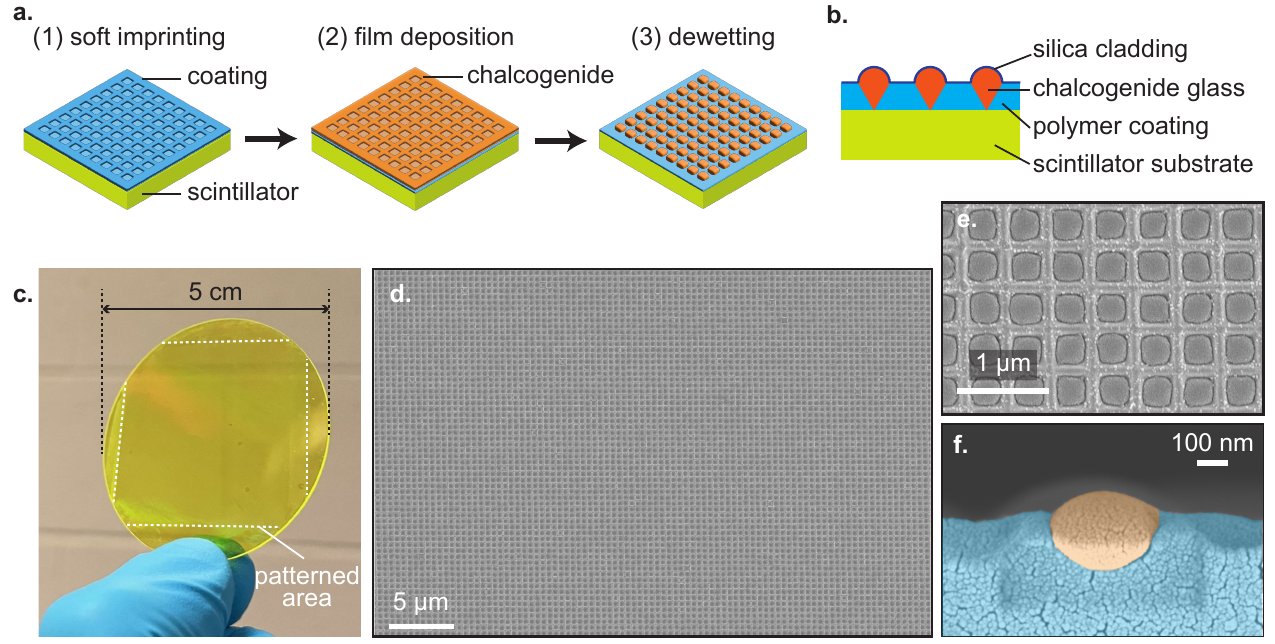}
    \caption{\small \textbf{Fabrication scheme of self-assembled nanophotonic scintillators.} \textbf{a.} Fabrication method: (1) soft nanoimprinting of polymer coating (blue) spin-coated over scintillator (yellow); (2) chalcogenide film deposition (orange); (3) annealing and dewetting. \textbf{b.} Cross-sectional schematic of nanophotonic scintillator. \textbf{c.} Photograph of large scale sample after fabrication of nanostructured layer on a  0.5 mm-thick YAG:Ce scintillator substrate (IL sample). \textbf{d.} Top view scanning electron micrograph of the dewetted nano-array (IL sample). \textbf{e.} Zoomed-in top view (IL sample). \textbf{f.} False color cross-sectional view of a single nanoparticle. From bottom to top, the cross-section shows the polymer coating (blue), a single chalcogenide nanoparticle (orange), and air.}
    \label{fig:fig1}
    \vspace{-0.3cm}
\end{figure*}

An emerging approach in scintillator research -- coined ``nanophotonic scintillators'' -- consists of structuring scintillator materials at the scale of their emission wavelength~\cite{roques2022, kurman2020} to control their emission properties, such as their light yield~\cite{roques2022, Knapitsch2014}, emission directionality~\cite{shultzman2023enhanced}, detection efficiency~\cite{ye2022enhancing}, and timing~\cite{ye2024nanoplasmonic}. Such enhancements open up new possibilities for more precise and efficient X-ray imaging technologies~\cite{roques2023free, singh2024bright, shultzman2024towards}.

% The integration of scintillators with nanophotonic structures, known as "nanophotonic scintillators," has shown significant promise in further enhancing scintillation performance. This technique involves embedding nanophotonic structures within scintillator materials to manipulate and control the emission of light at the nanoscale. In a notable study, this approach resulted in a 2.3-fold increase in scintillation signal over a 410 µm × 410 µm area of a relatively thick (>50 microns) scintillator patterned using focused ion beam milling~\cite{roques2022}. 

Despite these promising results, the widespread adoption of nanophotonic scintillators is hindered by challenges in scalable fabrication techniques. Current top-down fabrication methods, which rely on sophisticated lithographic techniques, offer nanometer-scale resolution and repeatability but are often complex, costly, difficult to scale to large areas, and not transferable to ``unconventional'' substrate materials such as scintillator crystals. On the other hand, bottom-up approaches such as laser printing ~\cite{zywietz2014, Urban2010}, chemical or self-assembly methods~\cite{fan2010,Vidgerman2012}, and topographical control of particle positioning ~\cite{flauraud2016} have been explored as alternatives. However, these methods come with their own set of limitations, including throughput constraints, surface defects, roughness, and restricted material choices, which also hinder their practical application. To maximize the technological impact of nanophotonic scintillators, it is imperative to develop
fabrication techniques that are scalable to industry-standard detector dimensions ($\sim$ cm), while preserving enhancements obtained from nanophotonic patterning.

% Current top-down fabrication methods, which rely on sophisticated lithographic techniques, offer high resolution and repeatability but are often complex, costly, and difficult to scale to large areas. On the other hand, bottom-up approaches such as laser printing ~\cite{zywietz2014, Urban2010}, chemical or self-assembly methods ~\cite{fan2010,Vidgerman2012}, and topographical control of particle positioning ~\cite{flauraud2016} have been explored as alternatives. However, these methods come with their own set of limitations, including throughput constraints, surface defects, roughness, and restricted material choices, which also hinder their practical application. To maximize the technological impact of nanophotonic scintillators, it is imperative to develop fabrication techniques that are scalable and compatible with large-area substrates, while preserving nanophotonic-enabled scintillation enhancements. 

Here, we demonstrate a large-scale nanophotonic scintillator fabrication method that realizes a six-fold nanophotonic enhancement in light yield over centimeter scales. Our method is based on the self assembly of chalcogenide glass photonic crystals (PhC). 
% Presently, top-down fabrication methods relying on lithographic techniques, offer high resolution and repeatability but are complex, costly, and challenging to scale to large areas. Alternatively, bottom-up approaches like laser printing~\cite{zywietz2014}, chemical or self-assembly methods~\cite{fan2010}, and topographical control of particle positioning~\cite{flauraud2016} have been explored, but come with their own limitations related to throughput, surface defects, roughness or limited material choices. 
With the devised method, we realize nanophotonic scintillators over an area of 4~cm~$\times$~4~cm (e.g., comparable in length scale to commercial flat panel detectors) with six-fold light yield enhancements compared to a reference bare scintillator. We demonstrate this enhancement in a conventional, and widely used X-ray scintillator material (cerium-doped yttrium aluminium garnet, YAG:Ce) and our method is in principle substrate-agnostic. We then elucidate the influence of surface disorder in our nanophotonic scintillator's imaging performance and obtain large-area scans of biological and inorganic specimens. Our results are poised for rapid integration into industrial applications, enabling a new family of optimized nanophotonic-enhanced scintillators for use in medicine, defense, and beyond.

\section{Nanophotonic scintillation enhancement over centimeter scales}

% Could start from the rationale above about the key properties: what are some key fabrication choices that enable them? How do they differ from the other previous methods (can

The depicted fabrication method seamlessly integrates a thin chalcogenide PhC coating atop a scintillating substrate (Fig.~\ref{fig:fig1}(a, b)). Chalcogenide glasses represent a particularly relevant class of materials for nanophotonic scintillation enhancement due their high refractive index ($2 \leq n \leq 4$) and low optical losses from the infrared to the visible spectrum~\cite{zakery2003optical}. Furthermore, thin (sub-100~nm) chalcogenide glass films exhibit viscous behavior during annealing over an extended processing window, making them ideal candidates for templated dewetting processes~\cite{das2019}. 

This process begins with the fabrication of a silicon master mold by traditional lithographic techniques (see Methods). Both interference lithography and electron beam lithography are used to make distinct molds, enabling different trade-offs between patterned area and resolution. The sample prepared with electron beam lithography extends over an area 4 mm × 4 mm (``EL'' in the rest of this work), while the sample prepared with interference lithography extends over a much larger area of 4 cm × 4 cm (``IL'' in the rest of this work). In the next process step, nanoimprint lithography is leveraged to reproduce the master texture onto the bulk scintillator. A polydimethylsiloxane stamp, replicated from a silicon master mold, is pressed onto a UV-curable polymer layer directly on the scintillator substrate and exposed to ultraviolet light (Fig.~\ref{fig:fig1}(a)-1). In a final process step, physical vapor deposition and dewetting is used to obtain a high-index nanoparticle array. A thin layer of chalcogenide glass is deposited using thermal evaporation (Fig.~\ref{fig:fig1}(a)-2). A final glass annealing step above its glass transition temperature induces the re-arrangement of the film into an array of highly ordered nanospheroids (Fig.~\ref{fig:fig1}(a)-3). The precise manipulation of interfacial tension as well as film-texture interaction is instrumental in achieving defect-free nanostructures, as exemplified in Fig.~\ref{fig:fig1}(d,e). The PhC exhibits a subwavelength period of 450 nm, covering a total patterned area of 4 cm × 4 cm, therefore counting around 10 billion nanoscale spheroids on the chip. Finally, a 15~nm SiO$_2$ layer is sputtered over the resulting chalcogenide nanoparticle array to avoid further oxidation.

\begin{figure}
\centering
\vspace{-0.2cm}
  \includegraphics[scale=0.7]{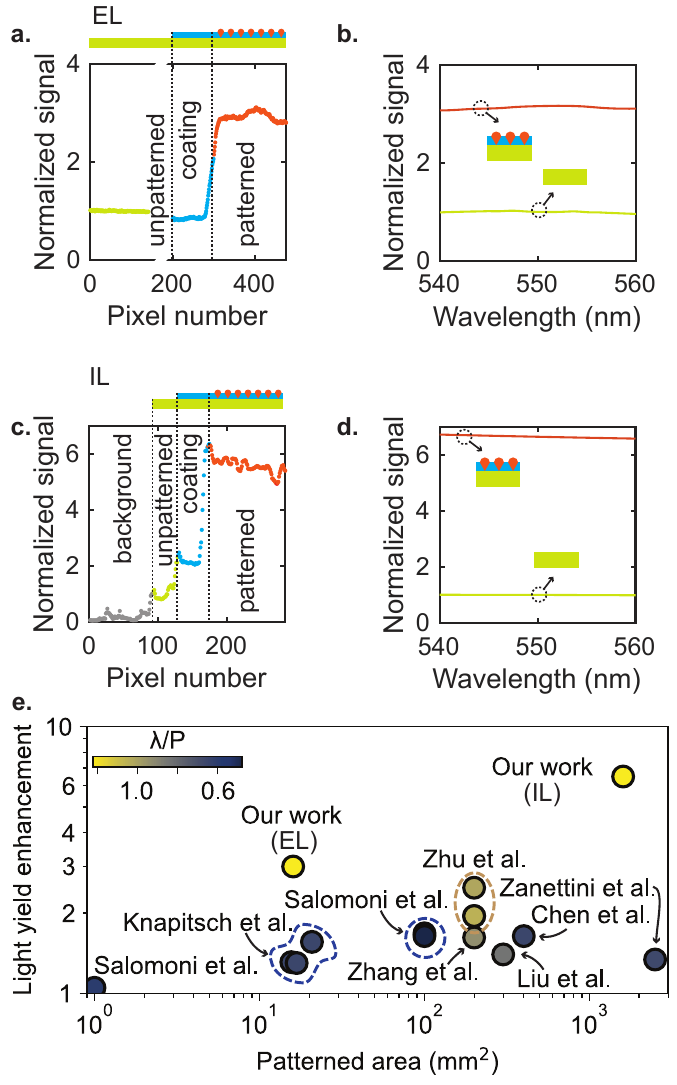}
    \caption{\small \textbf{Large-area nanophotonic enhancement of scintillation.} \textbf{a, c.} Intensity line scan (EL sample with $t_1=1$~mm (\textbf{a.}) and IL sample with $t_2=0.5$~mm (\textbf{c.})) under X-ray exposure across unpatterned, coated, and patterned areas of the sample. The coating refers to the polymer coating and silica cladding depicted in Fig.~
    \ref{fig:fig1}, without chalcogenide. The pattern refers to the polymer coating, chalcogenide nanoparticle, and silica cladding. \textbf{b, d.} Corresponding theoretical predictions for unpatterned and patterned scintillators. Data is normalized to the mean signal value from the unpatterned signal. \textbf{e.} Large-area nanophotonic scintillator benchmark: light yield enhancement and patterned area for all comparable devices reported in the literature. The dot color corresponds to the ratio of wavelength $\lambda$ to photonic crystal periodicity $P$. Subwavelength designs correspond to $\lambda/P>1$. Works appearing on this plot are Refs.~\cite{Salomoni2017, salomoni2018, Knapitsch2014,Zanettini2016, Liu2017, Zhang2017, Zhu2015, Zhu2015a, Zhu2015b, chen2018enhanced}.}
    \label{fig:fig2}
    \vspace{-0.3cm}
\end{figure}

Using a recent framework to model scintillation emission in nanophotonic structures, we anticipate an enhancement in nanophotonic scintillation by amplifying light yield due to better in/out-coupling of light (which maps to an enhancement in non-equilibrium optical absorption in the scintillator layer, via Lorentz reciprocity) ~\cite{roques2022}, as experimentally observed in Fig.~\ref{fig:fig2}(a, c). Taking the bare scintillator as reference (no coating nor pattern), our simulations indicate an increase in scintillation light yield of 3.48-fold (EL) and 6.96-fold (IL). These enhancement values are confirmed experimentally, with X-ray line scans showed in Fig.~\ref{fig:fig2}(c,d): $3.00 \pm 0.21$ (EL) and $6.62 \pm 1.6$ (IL).
Since the fabricated structures are spheroids, our numerical simulations consist of a multi-step process that combines finite element methods and rigorous coupled wave analysis~\cite{jin2020inverse} (see Methods).
% Moreover, the nanophotonic scintillator induces no obvious loss in resolution, as exemplified by imaging of a metallic grid in Fig. 1(f). 

We attribute the difference in nanophotonic scintillation enhancement between the two samples to several factors. First, we find in our simulations that thicker scintillators generally exhibit less nanophotonic enhancement, which is consistent with an analysis based on density of states~\cite{yu2010fundamental, benzaouia2024theory} (the samples' thickness are $t_1$ = 1~mm and $t_2$ = 0.5~mm). Second, we also observe experimentally the influence of different dewetting schemes and mold quality~\footnote{Compared to the bare scintillator, the coating on the EL sample (resp., IL sample) slightly reduces (resp. increases) the light out-coupling efficiency. This can be attributed to random pattern formation that occurs when dewetting flat surfaces~\cite{xue2011pattern}.}. 

Compared to previous works that aimed at realizing large-area micro or nanostructures on scintillators, our work realizes a six-fold nanophotonic enhancement over a patterned area of $>$1,600~mm$^2$ (which is only limited by the size of the available scintillator substrates). An overview of the state of the art is shown in Fig.~\ref{fig:fig2}(e). Previous work with comparably large patterned areas realized enhancements $\sim 1.3$.
We mainly attribute this five-fold improvement over the state of the art by the use of subwavelength nanophotonic structures. Periodic nanophotonic structures of period $P$ such that $\lambda/P \gtrsim 1$ are known to be optimal in terms of density of states enhancement~\cite{yu2010fundamental, roques2022}. The fact that our method is compatible with patterning of high index, subwavelength structures is therefore key in scaling up nanophotonic scintillator technology to areas required for X-ray imaging applications.

\section{Large-area X-ray imaging}

Using the larger area IL sample (4~cm $\times$ 4~cm) in the X-ray imaging setup shown in Fig.~\ref{fig:disorder}(a), we realized X-ray scans of inorganic and organic specimens. The X-ray imaging parameters (source voltage and power, as well as geometric and objective magnifications, can be found in the Methods and in the Supplementary Information (SI), Section~S4). Each X-ray scan is taken in conjunction with an X-ray flat field image used for post processing, and the final brightness and contrast are digitally adjusted (as would be done in a commercial X-ray scanner for industrial or medical applications).

\begin{figure}
\centering
\vspace{-0.2cm}
  \includegraphics[scale=0.65]{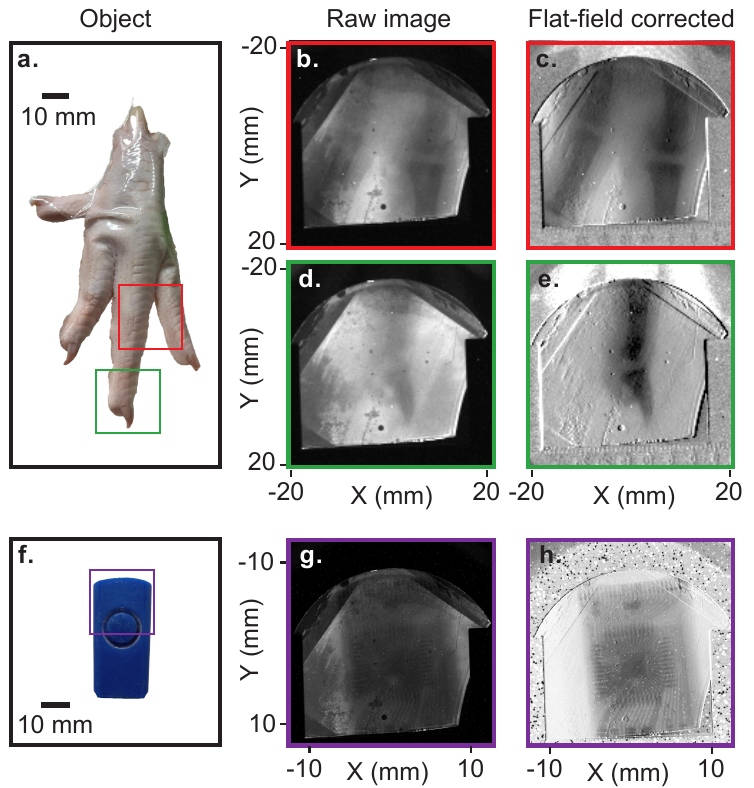}
    \caption{\small \textbf{X-ray imaging with large-area nanophotonic scintillators.} \textbf{a, f.} Photograph of different objects imaged through our customized imaging setup. The colored square denotes estimated field of views. \textbf{b, d, g.} Corresponding raw X-ray images. \textbf{c, e, h.} Corresponding flat-field corrected and contrast-adjusted X-ray images.}
    \label{fig:fig4}
    \vspace{-0.3cm}
\end{figure}

We first image parts of a chicken foot (tarsometatarsus and digits, shown in Fig.~\ref{fig:fig4}(a-e)). We clearly see several phalanges separated by interphalangeal joints: two proximal and middle phalanges in Fig.~\ref{fig:fig4}(b,c), and two intermediate and distal phalanges in Fig.~\ref{fig:fig4}(d,e). We then image a USB stick in Fig.~\ref{fig:fig4}(f-h) and can distinguish multiple levels of printed circuits overlayed on top of each other. These images demonstrate the potential of our scintillators to realize X-ray scans of centimeter-large objects with nanophotonic enhancement.

\section{Controlling spatial resolution with disorder in photonic crystal scintillators}

Next, we elucidate the influence of fabrication disorder on the spatial resolution of nanophotonic scintillators. 

\begin{figure}
\centering
\vspace{-0.2cm}
  \includegraphics[scale=0.58]{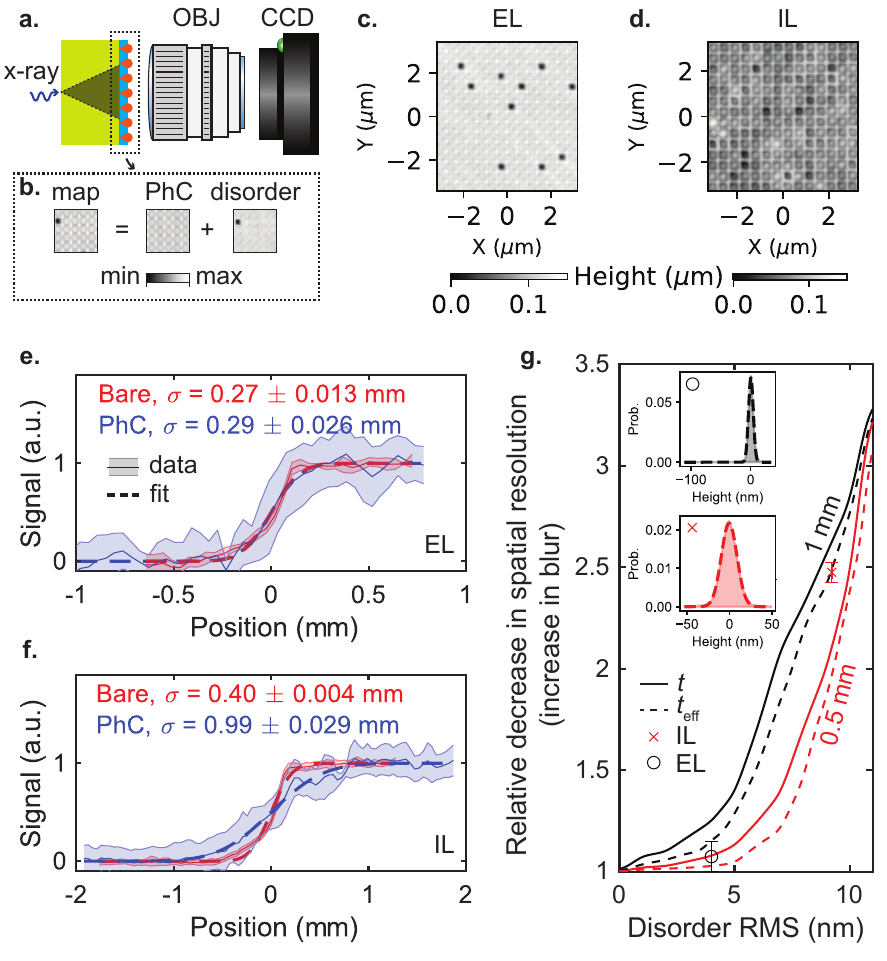}
    \caption{\small \textbf{Controlling spatial resolution with disorder in photonic crystal scintillators.} \textbf{a.} Experimental setup for X-ray imaging with large-area nanophotonic scintillators. \textbf{b.} The transfer function of the disordered photonic crystal is modeled as a convolution between that of an ordered photonic crystal (PhC) and that of a disordered height map.  \textbf{c, d.} Atomic force micrographs (AFM) for EL (\textbf{c.}) and IL samples (\textbf{d.}). \textbf{e, f.} Corresponding spatial resolution measurements in the presence of a razor blade to block part of the incoming X-rays. \textbf{g.} Relative decrease in spatial resolution (increase in blur) as a function of the disorder root mean square (RMS). Full lines are calculated for fields propagating through the whole scintillator (thickness $t$), dashed lines for its effective thickness $t_\text{eff}$ taking into account X-ray absorption. The dots represent measurements for the two samples (EL and IL) in this paper and insets show their respective disorder distribution from AFM.}
    \label{fig:disorder}
    \vspace{-0.3cm}
\end{figure}

When comparing atomic force microscopy (AFM, shown in Fig.~\ref{fig:disorder}(c,d)) images of the two samples, we observe various levels of disorder which we attribute to different lithography methods used to realize the nanoimprinting mold. A disorder distribution is extracted for both samples using Fourier analysis (see SI, Section~S3 and Fig.~\ref{fig:disorder}(b)). We also experimentally characterized each sample's spatial resolution by X-ray imaging the sharp edge of a razor blade (with a setup shown in Fig.~\ref{fig:disorder}(a)). 

Generally, we observe a correlation between greater amounts of nanofabrication disorder and a decrease in the scintillator's spatial resolution (blur increase). Specifically, the nanofabricated pattern on the EL sample (Fig.~\ref{fig:disorder}(e)) has no significant influence on its spatial resolution (with a slight relative decrease within experimental uncertainty). However, the IL sample (Fig.~\ref{fig:disorder}(f)) exhibits a decrease in relative spatial resolution by a factor of of $2.48 \pm 0.10$. Dependence on the X-ray energy of the spatial resolution is analyzed in the SI, Section~S5. 
% A limitation of our measurement of spatial resolution is that it is not necessarily consistent with the intuition that thicker scintillators have lower spatial resolution due to greater optical blur on their back facet. In our measurements, the thicker sample (EL) appears to have better spatial resolution than the thinner one (IL). This is because our imaging system was focused by hand to achieve the sharpest possible image, without precisely tracking the position of the focal plane. Therefore, it is possible that the focal plane when imaging the EL sample was closer to the front (patterned) facet than the case of the IL sample, resulting in, effectively, lower blur. In the future, this can be mitigated by either tracking the focal plane or keeping it constant between measurements.
A limitation of our measurement of spatial resolution is that our imaging system was focused by hand to achieve the sharpest possible image, without precisely tracking the position of the focal plane. Therefore, it is possible that the focal plane when imaging the EL sample was closer to the front (patterned) facet than the case of the IL sample, resulting in, effectively, lower blur. In the future, this can be mitigated by either tracking the focal plane or keeping it constant between measurements.
% We also note that our measurement of spatial resolution is consistent with the intuition that thicker scintillators have lower spatial resolution due to greater optical blur on their back facet (here, by about a factor of 1.5). 

To account for the influence of disorder on spatial resolution, we adapted the framework of stochastic surface transfer functions~\cite{harvey2013parametric} to X-ray scintillation imaging. The disorder distribution from AFM measurements is modeled as a Gaussian-distributed surface transfer function that blurs optical waves incident from within the scintillator. We calculated the relative decrease in spatial resolution for both samples (corresponding to an increase in optical blur, shown in Fig.~\ref{fig:disorder}(g)), with scintillation emission happening on the front facet of the scintillator (propagation through thickness $t$) and at the mean X-ray absorption position (propagation through thickness $t_\text{eff}$). The relative decrease in resolution is defined as the ratio of full widths at half maxima of the line spread functions (for a disordered photonic crystal vs. a flat scintillator surface) -- a value $>1$ corresponding to a decrease in resolution, greater blur, and coarser features of the X-ray scan. These \textit{ab initio} disorder simulations agree within 7\% (comparing data from the EL sample and simulations for $t_\text{1, eff}$) and 15\% (comparing data from the IL sample and simulations for $t_\text{2}$).
The remaining discrepancy originates from the uncertainty in the depth at which the imaging objective is focused. More details on the spatial resolution measurement method, disorder modeling, AFM analysis, and calculation of the effective thickness, can be found in the SI, Section~S2-3.

\section{Discussion} 

% These results draw some exciting perspective for the development of brighter and thinner X-ray scintillators which could potentially lead to low-dose and high-resolution X-ray imaging, with promising applications in medical imaging and nondestructive inspection.

We have presented a platform for nanophotonic scintillation that combines the following key features: (1) compatibility with $>$ centimeter-scale fabrication methods, to enable large-area X-ray imaging; (2) subwavelength feature sizes, to maximize nanophotonic outcoupling~\cite{roques2022, yu2010fundamental}; (3) absence of residual layer, which may reduce nanophotonic enhancement due to impedance mismatch~\cite{choi2023realization}; (4) ability to realize low-loss high-index ($n>2$) nanostructures, for greater nanophotonic control~\cite{das2019, khorasaninejad2016metalenses}; (5) high repeatability since a master mold can be used to generate thousands or tens of thousands of large-area nanophotonic scintillators~\cite{guo2007nanoimprint, chou1996nanoimprint}. %How about high repeatability? Conclusion form this paper
The combination of these factors allowed us to demonstrate scintillation enhancement over scales commensurable with that of commercial X-ray flat panel detectors and provides a path towards their mass production.

% For the first time, subwavelength, large-area, high index fabrication method -- resulting in almost an order of magnitude enhancement and on par with largest areas ever fabricated . This is enabled by ... mention some nice properties of the fab and materials (chalcogenides) 

We have also established a clear correlation between nanofabrication disorder and spatial resolution decrease. Improvements in mold quality realized with interference lithography will lead to lower amounts of nanofabrication disorder and little to no decrease in the scintillator's spatial resolution, while improving scintillation light yield six-fold.

Our work is also a first step towards the realization of large-scale metaoptics on scintillators, since the realized method is amenable to local control of the spheroid's shape and induced phase shift~\cite{das2019}. Proper design and optimization of metasurface masks could lead to coincident enhancements in spatial resolution and light yield~\cite{shultzman2023enhanced}, as well as local control the surface's transfer function~\cite{khorasaninejad2016metalenses, yu2011light, yu2014flat, overvig2021thermal}. 

Furthermore, the inherent scalability of our fabrication technique opens the door to patterning entire rolls of scintillator materials, paving the way for rapid and affordable mass manufacturing of nanophotonic scintillators. By enabling nanoscale control over electromagnetic properties with controllable disorder, our method ensures both high performance and robustness in practical applications. These advancements hold the potential to advance X-ray imaging, as they allow for the integration of complex nanophotonic structures over large areas without compromising quality or significantly increasing costs. 
These improvements may facilitate the wider adoption of nanophotonic scintillators in various fields requiring high-resolution, high-sensitivity detection of X-rays and other high-energy particles.

\section{Methods}

\textbf{Templated dewetting of chalcogenide on scintillator materials.} \textit{Mold Fabrication}: Two separate molds are prepared for nanoimprinting. A first mold (Mold 1) is prepared by electron beam lithography over an area of 4 mm × 4 mm. Following oxygen plasma cleaning and HMDS monolayer deposition, a negative electron beam resist (maN 2403, MicroResist Technology GmbH, Germany) is spun at \SI{3000}rpm on a silicon wafer covered with a $\SI{30}{\nano\meter}$ native oxide. The resist is soft baked at $\SI{90}{\degreeCelsius}$ for $\SI{2}{\minute}$. An inverted grid pattern is written using a \SI{50}{\keV} electron beam and $\SI{10}{\nano\ampere}$ current. The resist is developed using a TMAH-based developer (AZ-726, MicroChemicals GmbH, Germany) for $\SI{2}{\minute}$. The native oxide is selectively etched using ICP reactive ion etching (RIE 230iP, Samco, Japan) with fluorine chemistry.
A second mold (Mold 2) is prepared by interference lithography over an area of $\SI{40}{\milli\meter} \times \SI{40}{\milli\meter}$. A HeCd laser source (\(\lambda = \SI{325}{\nano\meter}\)) is directed at a pinhole placed approximately $\SI{60}{\centi\meter}$ away from a Lloyd's mirror setup. An antireflective coating layer is spun at \SI{3500} rpm onto a silicon wafer with a $\SI{30}{\nano\meter}$ native oxide (AZ Barli II 90, iMicroMaterials, Germany), and baked at \SI{180}{\degreeCelsius} for \SI{1}{\minute}. A layer of positive photoresist (AZ 3312, iMicroMaterials, Germany) is spun at \SI{5000} rpm over the antireflective coating, followed by a soft bake at \SI{110}{\degreeCelsius} for \SI{1}{\minute}. The second mold is exposed with an MLA 150 Advanced Maskless Aligner (Heidelberg Instruments, Germany) at \SI{135}{\micro\coulomb/\centi\meter\squared}, followed by a post-exposure bake at \SI{110}{\degreeCelsius} for \SI{2}{\minute}. The photoresist is developed in AZ 726 for \SI{2}{\minute}. The native oxide and antireflective coating are etched using ICP reactive ion etching (RIE 230iP, Samco, Japan) with fluorine chemistry. Both resulting molds are then stripped with an oxygen plasma cleaning step (e3511 Plasma Asher, ESI, USA). The resulting silicon wafer is placed in a $\SI{25}{\percent}$ KOH solution at $\SI{60}{\degreeCelsius}$ for anisotropic silicon etching for $\SI{2}{\minute}$. The native oxide hard mask is stripped with a $\SI{1}{\minute}$ dip in diluted HF $\text{10:1}$.

\textit{Nanoimprint Lithography}: The resulting silicon mold is treated with an anti-sticking layer (Trichloro(1H,1H,2H,2H-perfluorooctyl)silane, Millipore Sigma, USA) in a vacuum desiccator following a short oxygen plasma surface activation. A (poly)dimethylsiloxane layer (PDMS Sylgard 184, Corning, USA) is drop-casted on the silanized mold and cured at $\SI{80}{\degreeCelsius}$ for $\SI{2}{\hour}$. Upon curing, the PDMS layer is peeled off from the mold.

\textit{Scintillator Patterning}: A thin layer of diluted UV-Curable polymer (Ormocer, MicroResist Technology, Germany) is spun onto the YAG:Ce bulk scintillator. The PDMS nanoimprint mold is pressed directly onto the thin polymer layer. A UV light source (\(\lambda = \SI{375}{\nano\meter}\)) is shone through the PDMS to cure the polymer film, with a dose $>\SI{1500}{\milli\joule/\centi\meter\squared}$. A sub-$\SI{100}{\nano\meter}$ thin chalcogenide layer is deposited by thermal evaporation (PVD Products, USA) directly onto the patterned polymer film. The chalcogenide is annealed above its glass transition temperature to dewet according to the underlying texture. A $\SI{15}{\nano\meter}$ SiO$_2$ layer is sputtered over the resulting chalcogenide nanoparticle array to protect it from the surrounding oxidative environment (ATC Sputtering System, AJA International, USA).

\textbf{X-ray imaging experiments.} All experiments, including imaging and measurements of scintillation enhancement and spatial resolution, were done using a custom-built experimental setup inside of a ZEISS Xradia Versa 620 micro-CT. Images were captured using a Hamamatsu ORCA-Fusion C14440-20UP CMOS camera along with a wide-field-of-view camera lens (Edmund Optics 33-304). A narrow bandpass filter centered at 550 nm (AVR Optics, 15~nm bandwidth) was placed in front of the camera lens to minimize unwanted background from other wavelengths. In all experiments, the source was $d_{s}=150$ mm away from the scintillator, while the object distance $d_o$ depended on the desired geometric magnification of each image, defined as $M_{g}=d_{s}/d_{o}$. Here, we summarize the most important details of each experiment; further details of the experimental setup, methods, underlying theory, and data processing can be found in the SI.

\textit{Scintillation enhancement}: Scintillation enhancement was measured by directly imaging the scintillator under excitation from X-rays. The line profiles shown in Fig. \ref{fig:fig2} were extracted from these images. The EL sample was excited by X-rays at 60 kVp and 6.5 W, while the IL sample was excited by X-rays at 150 kVp and 23 W. The scintillation enhancement is calculated using the formula $(I_P-I_B)/(I_U-I_B)$, where $I_P$, $I_U$, $I_B$ are the average patterned, unpatterned, and background intensities from the line profile. 

\textit{Imaging}: To capture an image, an object was placed between the source and the scintillator. For each object, the focal plane---usually somewhere inside the scintillator rather than coplanar with the patterned surface---was adjusted by slightly moving the camera lens back and forth to capture the sharpest image. Brightness and contrast were adjusted by carefully tuning the X-ray energy (i.e., kVp), exposure time, and pixel binning and subarray. Figs. \ref{fig:fig4}(b) and \ref{fig:fig4}(d) were captured under the following conditions: X-ray energy and power of 30 kVp and 2 W, exposure time of 10 s, binning of 4, and subarray of 144 × 144, and geometric magnification of 1.2. Finally, Fig. \ref{fig:fig4}(g) was captured under the following conditions: X-ray energy and power of 60 kVp and 6.5 W, exposure time of 7 s, binning of 1, subarray of 576 × 576, and geometric magnification of 2. Post-processing included flat-field correction and digitally adjusting brightness and contrast (Figs. \ref{fig:fig4}(c), \ref{fig:fig4}(e), and \ref{fig:fig4}(h)). The flat-fields were captured by simply removing the objects without changing any of the aforementioned parameters. The chicken foot used in this study was obtained from a local grocery store and prepared for X-ray imaging following standard procedures, including vacuum sealing in a plastic bag to prevent contamination. The USB stick was a 3.0 USB Flash Drive Pen from OneSquareCore. 

\textit{Spatial resolution}: The spatial resolutions of the nanophotonic scintillators were estimated by imaging a carbon steel razor blade. Ideally, a region of interest drawn across the edge of the razor blade looks like a blurred, two-dimensional step function that can be reasonably approximated by an edge spread function (ESF) of the form $\textrm{ESF}(x,y)=(A/2)\textrm{erf}\left[(x-\mu)/\sqrt{2}\sigma\right]$, where the spatial resolution is $2\sqrt{2\ln 2}\sigma \approx 2.3548\sigma$. By fitting the raw data (not the flat-field corrected data, which may remove the effects of blurring due to thickness), $\sigma$ can be estimated. In Fig. \ref{fig:disorder}(e)--(f), the measured ESFs were captured under the following conditions: X-ray energy and power of 90 kVp and 12 W, exposure time of 3 s, binning of 1, and geometric magnification of 2. The EL sample (Fig. \ref{fig:disorder}(e)) used a subarray of 144 × 144, while the IL sample (Fig. \ref{fig:disorder}(f)) used a subarray of 576 × 576.

\textbf{Modeling nanophotonic enhancement.}  
\textit{Finite element + rigorous coupled wave analysis modeling}:
The nanophotonic scintillation enhancement is modeled using a three-step approach. The large thickness ($\sim$ mm) of the scintillator and the complex geometry of the nanophotonic structure make it challenging to use a single computational tool with high efficiency and low computational cost. Therefore, a combination of finite element (FE) and rigorous coupled wave analysis (RCWA) methods is employed. The nanophotonic scintillator consists of a chalcogenide spheroid with a diameter of about 395 nm, a subwavelength period of 450 nm, a polymer coating of 450 nm thickness, and scintillator substrate. The refractive index of SiO$_{2}$ is obtained from literature~\cite{rodriguez2016self}, while that of the Ormocer polymer (1.5) and YAG:Ce are given by the suppliers. The refractive index of chalcogenide is obtained from in-house ellipsometry measurements (see SI, Fig.~S2). All calculations were carried out in the wavelength range of 540 to 560 nm, with the YAG:Ce emission peak centered at 550 nm.

We utilize a commercially available FE solver, COMSOL Multiphysics\textregistered, to model the electromagnetic response of the spheroid nanophotonic structure. Initially, we simulate the superstrate without YAG:Ce (polymer coating and chalcogenide spheroid), then combine it with a thinner (1~µm) YAG:Ce scintillator to optimize the geometry. Next, we replace the spheroids with multiple stacked cylinders that approximate the electromagnetic response (transmission and reflection) of the full spheroid structure calculated in the first step (still with an FE solver). This is so that we can shift from FE simulations -- where simulating a thick substrate would be difficult to model -- to RCWA simulations, where we can use our approximated structure and a thick substrate since it is a semianalytical method. In the third step, we use an automatically differentiable RCWA solver~\cite{jin2020inverse}, to simulate the approximate geometry with the thick substrate. The absorption/emission (via reciprocity) within/from the volume of the scintillator is calculated for both polarizations (transverse electric (TE) and transverse magnetic (TM)) and averaged to mimic unpolarized light. Finally, we calculate the enhancement factor as the ratio of the spectrally integrated emission of the patterned scintillator to the unpatterned scintillator.

\textit{Influence of disorder on spatial resolution:} The image degradation due to disorder within the nanophotonic structure is analyzed using the surface scatter theory based on the linear shift-invariant system formulation~\cite{harvey2020retrospective}. This is executed in several steps, following traditional image formation theory. First, the modulation transfer function (MTF) of the nanophotonic spheroids is calculated using the transfer function retrieved directly from the transmission as a function of angle (using RCWA). Consequently, the disorder is modeled using the surface scatter theory to calculate the surface transfer function (STF). The scattering property by the disorder (defects) is considered in transmission mode to formulate the STF (see SI, Section S2). Here, we utilize a Gaussian autocovariance function with surface roughness root mean square (RMS) retrieved from AFM images using Fourier analysis (see SI, Section S3). 

Further, we define the system modulation transfer function (MTFsys), which incorporates the influence of nanophotonic spheroids and disorder, via the two transfer functions: MTF and STF (multiplication in the Fourier space). MTFsys provides a complete linear system formulation of image quality as degraded by surface scatter effects due to disorder from residual optical fabrication errors. Once the MTFsys is derived, we proceed to obtain the line spread function (LSF) through an inverse Fourier transform to real space. This is followed by propagation through the optical imaging setup, encompassing free space and optical components. The full width at half maximum (FWHM) of the LSF serves as the defining measure of the system's resolution. These procedures are consistently applied to both EL and IL samples in both effective (0.76, 0.36 mm) and real (1, 0.5 mm) thicknesses (see definition of the effective thickness in SI, Section S2). Subsequently, the FWHM of the nanophotonic structure is compared with that of a smooth unpatterned scintillator to calculate the relative spatial resolution as a function of RMS (1 – 11 nm). The calculations above are performed by considering the peak emission wavelength (550 nm) of the scintillator.

\section{Authors contributions}
L.~M.-M. and C.~R.-C. conceived the original idea. 
L.~M.-M. developed the fabrication methods.
S.~P. built the experimental setup and acquired the data with contributions from J.~C., L.~M.-M., S.~V., S.~E.~K., and C.~R.-C.
Experimental data and processing were performed by S.~P., M.~G.~A., and C.~R.-C.
Sample characterization (scanning electron and atomic force micrographs) were obtained by L.~M.-M. and S.~E.~K.~
M.~G.~A. performed the numerical simulations with contributions from L.~M.-M., S.~P., J.~C., S.~V., S.~M., S.~C., and C.~R.-C.~
C.~R.~C., M.~S., J.~H., and B.~M. supervised the project. 
The manuscript was written by L.~M.-M., S.~P., M.~G.~A., and C.~R.-C. with inputs from all authors.

\section{Competing interests}
The authors declare no potential competing financial interests.

\section{Data and code availability statement}
The data and codes that support the plots within this paper and other findings of this study are available from the corresponding authors upon reasonable request. Correspondence and requests for materials should be addressed to L.~M.~M. (lmmartin@mit.edu), S.~P. (pajovics@mit.edu), M.~G.~A. (abebe181@mit.edu), and C.~R.-C. (chrc@stanford.edu).

\section{Acknowledgements}
The authors would like to thank Paul Lecoq, Ido Kaminer, Irina Shestakova, Olivier Philip, Bipin Singh, and Vivek Nagarkar for stimulating discussions. 
The authors thank Tim Savas for advice on interference lithography and Chris Hogan for experimental assistance.
This work is supported in part by the DARPA Agreement No.~HO0011249049, as well as being also supported in
part by the U. S. Army Research Office through the Institute for
Soldier Nanotechnologies at MIT, under Collaborative Agreement Number W911NF-23-2-0121.
L. M. M. was supported by the DARPA ENvision program and a fellowship from the Swiss National Science Foundation (P500PT-203222). 
S. P. gratefully acknowledges support from the NSF GRFP under Grant No. 2141064.
M. G. A. is funded by the Fonds de la Recherche Scientifique - FNRS (Grant No.  FC 053809) and Wallonie-Bruxelles International (WBI) fellowship.
C.~R.-C. is supported by a Stanford Science Fellowship. 

\bibliographystyle{ieeetr}
\bibliography{bibliography}

\end{document}